\documentclass[iop,revtex4,twocolappendix]{emulateapj}
\newcommand{\ifpp}[1]{#1}
\newcommand{\ifms}[1]{}
\ifpp{\usepackage{epsfig}}

\usepackage{ifpdf}
\usepackage{ifthen}
\usepackage{pdflscape}
\usepackage[pdftex]{hyperref}
\usepackage{pslatex}
\usepackage{url}

\begin{document}
\lefthead{Bowyer et al.}
\righthead{SERENDIP III}

\title{The SERENDIP III 70 cm Search for Extraterrestrial Intelligence}
\author{Stuart Bowyer, Michael Lampton, Eric Korpela, Jeff Cobb, Matt Lebofsky, and Dan Werthimer}
\affil{Space Sciences Laboratory, University of California, Berkeley, CA 94720}

\begin{abstract}
We employed the SERENDIP III system with the Arecibo radio telescope to
search  for possible artificial	extraterrestrial 
signals.  Over the four years of  this search we covered 93\% of the sky
observable at Arecibo at least once  and 44\% of the sky five times or 
more  with a sensitivity of $\sim 3\times 10^{-25} Wm^{-2}$.  
The data were sent to a $4\times 10^6$ channel spectrum analyzer.
Information was obtained from over $10^{14}$ independent data points
and the results  were then 
analyzed via a suite of pattern detection algorithms to identify narrow band 
spectral power peaks that were not readily identifiable as
the product of human activity.
We separately selected data coincident with
interesting nearby G dwarf stars that 
were encountered by chance in our sky survey for suggestions of excess
power peaks.  
The peak power distributions in both these data sets were
consistent with random noise. 
We report upper limits on possible signals from the stars investigated
and provide examples of the  most interesting candidates identified in the sky 
survey.
{\em This paper was intended for publication in 2000 and is presented here
without change from the version submitted to ApJS in 2000.}
\end{abstract}

\section{Introduction}

Early radio searches for extraterrestrial intelligence  used 
dedicated telescope time to search for emission from nearby stars (see 
Tarter 1991 for a partial listing, and Tarter 2001
for a full listing of these searches).
This type of search became increasingly 
difficult to carry out at major facilities
because of a general reluctance to devote dedicated telescope 
time to such projects, which though interesting, are acknowledged 
to have a low probability of success.  In addition, sky surveys were 
carried out which scanned substantial portions of the sky.   
The Berkeley SERENDIP project 
(Search for Extraterrestrial Radio Emissions from Nearby Developed 
Intelligent Populations) solved the dedicated telescope time problem by 
using data obtained simultaneously with ongoing astronomical research.  
This program began over twenty years ago (Bowyer et al. 1983) and has 
continued to the present day with ever increasing sensitivity
and an ever-widening set of search parameters.

Other sky surveys using dedicated telescopes have been, and are 
continuing to be, carried out.
The Ohio State 
program (Dixon 1985) was the earliest sky survey; this project 
is now terminated.
The Harvard search (Leigh \& Horowitz 1997) has also been terminated.
The Argentinian search (Lemarchand et al. 1997), and the Australian search
(Stootman et al. 1999) continue.   
Targeted searches of nearby stars have been initiated by the SETI 
Institute (Tarter 1997) using substantial amounts of dedicated telescope time
which were obtained in return for a substantial financial contribution to the
telescope upgrade which was carried out after the conclusion of the 
SERENDIP III observations.
	
We discuss our sky survey search for artificial extraterrestrial signals with the 
SERENDIP III system
(Bowyer et al. 1997) and the Arecibo telescope.
Although the search was a sky survey, nearby solar-type stars
inevitably fell within the beam pattern of the telescope in the course
of these observations.   As part of the analysis of the SERENDIP
III data, we have separately investigated the data from observations of the
sky coincident with these nearby stars.  Although our
integration times for individual targets are relatively short (as
compared, for example, with the SETI  Institute targeted search), our sensitivity is
still substantial because of the large collecting area of the Arecibo
telescope and the outstanding receivers that are available for use with
this instrument.  

We report the results of our sky survey and provide upper limits on possible
signals from stars in this paper.

\section{Observations and Analysis}

The SERENDIP III data were obtained with the National Astronomy
Ionosphere Center's radio observatory in Arecibo, Puerto Rico with a
430 MHz receiver.  
Data collection
began in April, 1992 and ended in October, 1998.  

The feed used for SERENDIP III was located at the opposite carriage house
from the primary observer's feed.  This resulted in three observation modes.
In the first, representing about 45\% of the observing time, the primary 
observer's feed was tracking a point on the sky.  This resulted in the 
SERENDIP III beam moving across the sky at roughly twice the sidereal rate.  
In the second, representing about 40\% of the observing time the feeds were 
stationary, resulting in motion at the sidereal rate.  In the third, 
representing about 15\% of the observing time, the SERENDIP III feed was
tracking a point on the sky.  
\ifpp{
\begin{figure*}[tbp]
\begin{center}
\includegraphics[width=\textwidth,keepaspectratio=true]{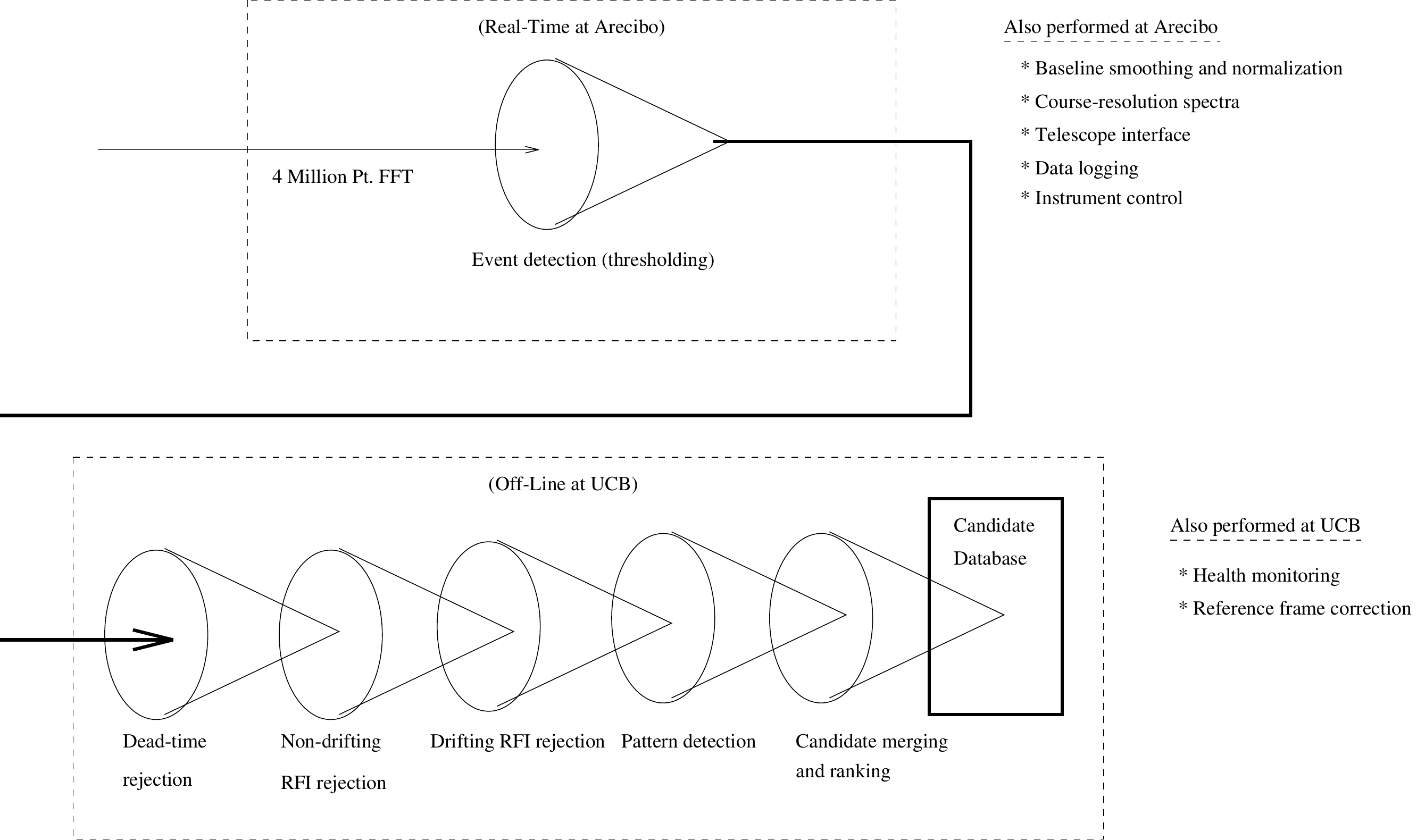}
\end{center}
\caption{An illustration for the SERENDIP III data flow showing the steps taken
to detect candidate signals.
Real-time processing consists of power spectra generation
and application of an adaptive threshold to the resulting spectra.  Off-line
data reduction included RFI rejection techniques, pattern detection,
and candidate extraction.\label{dataflow}}
\end{figure*}
}

Power spectra were generated by the SERENDIP III four million
channel spectrum analyzer which
had a 1.7 second integration period and a 0.6 Hz
frequency resolution (2.5 MHz instantaneous band coverage). The sensitivity
achieved in a 1.7 second integration at our typical system temperature
of 45 K was $3 \times 10^{-25}$ W m$^{-2}$. This bin
size is wide enough to encompass Doppler frequency drifts caused by the
Earth's motions plus reasonable accelerations of a hypothetical
transmitter's reference frame.  The receiver's entire
12 MHz band was processed in 2.4
MHz steps taking about 8.5 seconds to complete a single sweep. The
SERENDIP III fast Fourier transform based
hardware is described in detail by Werthimer, et al. (1997).

Adaptive thresholding was achieved by baseline smoothing
the raw power spectra with a sliding eight thousand channel local-mean boxcar
and searching for channels exceeding 16 times the mean spectral power.
These signals were recorded along with time, pointing
coordinates, detection frequency, and signal power.

Following the real-time data analysis by
the SERENDIP III instrument, the reduced data was shipped to Berkeley
of off-line data reduction and candidate 
generation.  
Off-line data reduction included 
radio frequency interference (RFI)
rejection techniques, pattern detection,
and candidate extraction.  The overall data flow is shown in Figure 
\ref{dataflow}.

\subsection{Off-line Data Reduction}

Data were transferred from Arecibo to Berkeley across the
Internet where off-line data analysis activities began.
Data reduction consisted of removing data taken during
periods when the telescope was slewing too fast or too slow
for our analysis procedures,
followed by the application of a suite of RFI filters.

Excessively rapid telescope slew rates
precluded acquisition of accurate
positioning information.  In addition, during times when the receiver tracks
a point on the sky,
it is not possible for our analysis programs
to differentiate between continuous RFI and a potential signal.
Therefore, our first filter was to censor data acquired during periods of rapid
telescope movement and periods when the telescope was tracking sky objects.
Roughly 15\% of the data were removed for this reason.

The next step in data reduction was non-drifting RFI rejection.
SERENDIP's non-drifting RFI rejection algorithms incorporated
dynamically adaptive statistical analysis
routines
that detect spurious signals
from terrestrial and near-space sources.
Three cluster analysis tests
were conducted on each input data file spanning several hours of
observation.
Signals were rejected if they
(1) were detected over broad areas
of the spectrum in one or more integration periods
(broad spectrum interference),
(2) persisted at the same receiver frequency through
multiple telescope beams, or
(3) persisted in the same channel of the spectrum analyzer.
 
Broad spectrum interference was identified by the 
rejection test:
 
\begin{equation}
x > 50
\hspace{.1in}
and
\hspace{.1in}
\frac{\sum{d^2}}{x} > S
\end{equation}
 
\noindent
where {\it d} is the number of frequency bins (0.6 Hz/bin) between simultaneous
events above the threshold and $x$ is the
number of events above threshold in the spectrum.
For SERENDIP III, the threshold value
was set at {\it S} = $10^8$.
 
RFI rejection algorithm (2) uses a statistical method to
determine if several detections at the same observing
frequency could be ruled out.  If these detections occur
with a significantly above-average hit rate, and they
continue to occur when our observing beam has moved beyond
one beam width (0.17 degrees), we reject the hypothesis of random Poisson
events being the cause.  Instead we mark these detections
as being due to external RFI, and reject them.

RFI rejection algorithm (3) uses the same statistical test
as (2) but applies it to hit sequences that have the
same intermediate-frequency bin number.  In this way it rejects
interference generated within the observatory.

Data surviving the first three rejection
criteria were further analyzed for RFI that
drifts rapidly in frequency
and were therefore not rejected by algorithms (2) and (3) above.
Figure \ref{rfirej}\ illustrates SERENDIP's drifting frequency RFI detection
algorithm.
\ifpp{
\begin{figure}[btp]
\begin{center}
\includegraphics[width=\columnwidth,keepaspectratio=true]{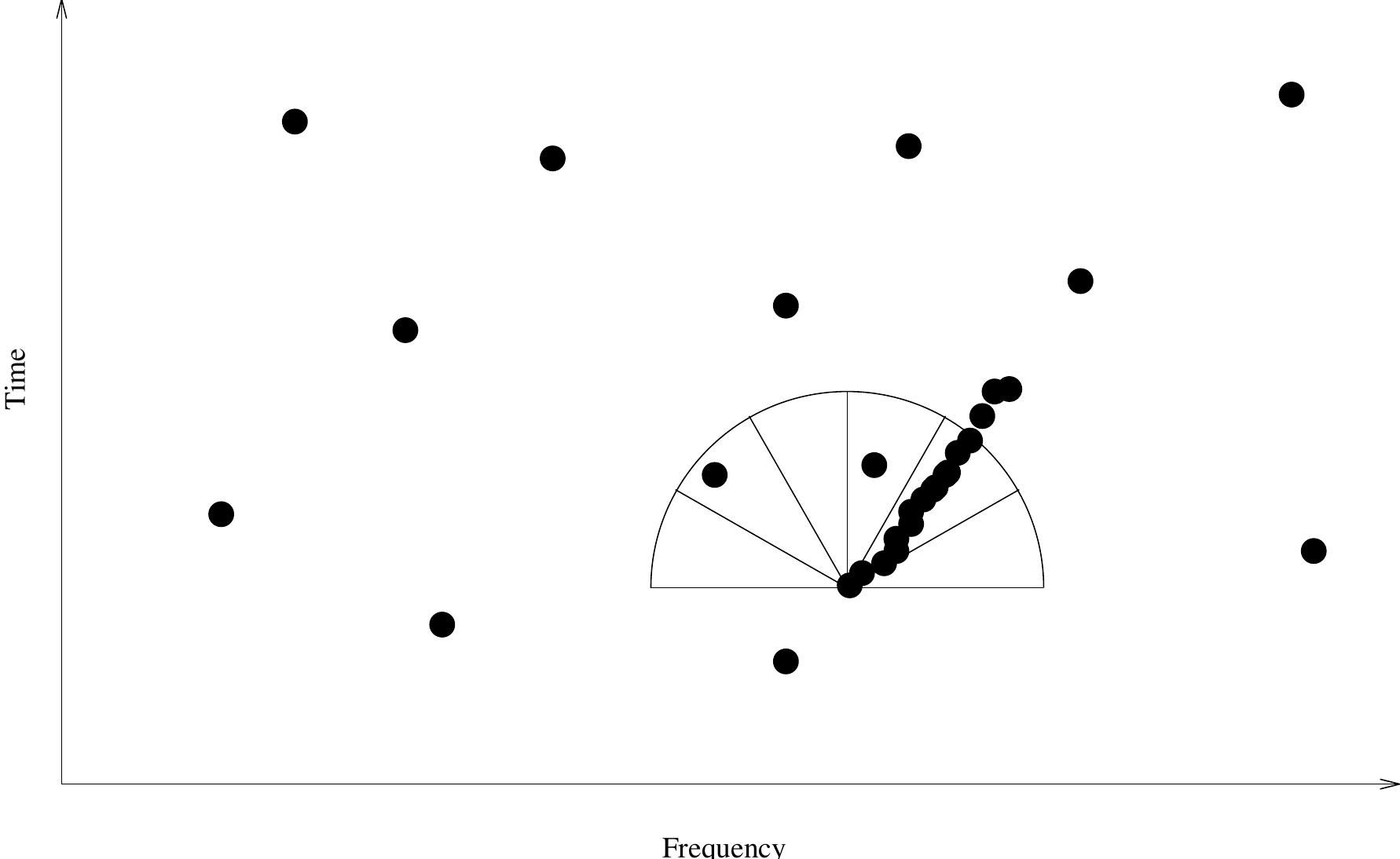}
\caption{Schematic illustration  of SERENDIP's drifting RFI detection algorithm.
Sectors of frequency-time space around each detection
were analyzed for signals that exhibit time-coherency
and persistence over multiple pointings of the telescope's beam.\label{rfirej}}
\end{center}
\end{figure}
}
The average event density in frequency-time space
was calculated for each data set.
Sectors in frequency-time space around each detection
(shown as dots in Figure 2) were
analyzed to identify sectors having an unusually high
number of detections.
Sectors containing an excessive number of detections
were further analyzed for
the presence of drifting signals that
persisted through multiple pointings of
the telescope beam.
 
An example of the efficacy of the RFI rejection techniques is shown in
Figure \ref{rfiexample}.
Note that after
application of SERENDIP's RFI filters, the remaining data closely approached the
number expected from a Poisson noise distribution.
Typically, 98\% of the RFI was rejected, yet only 1\% of the band
was lost during RFI removal.
\ifpp{
\begin{figure}[tbp]
\begin{center}
\includegraphics[width=\columnwidth,keepaspectratio=true]{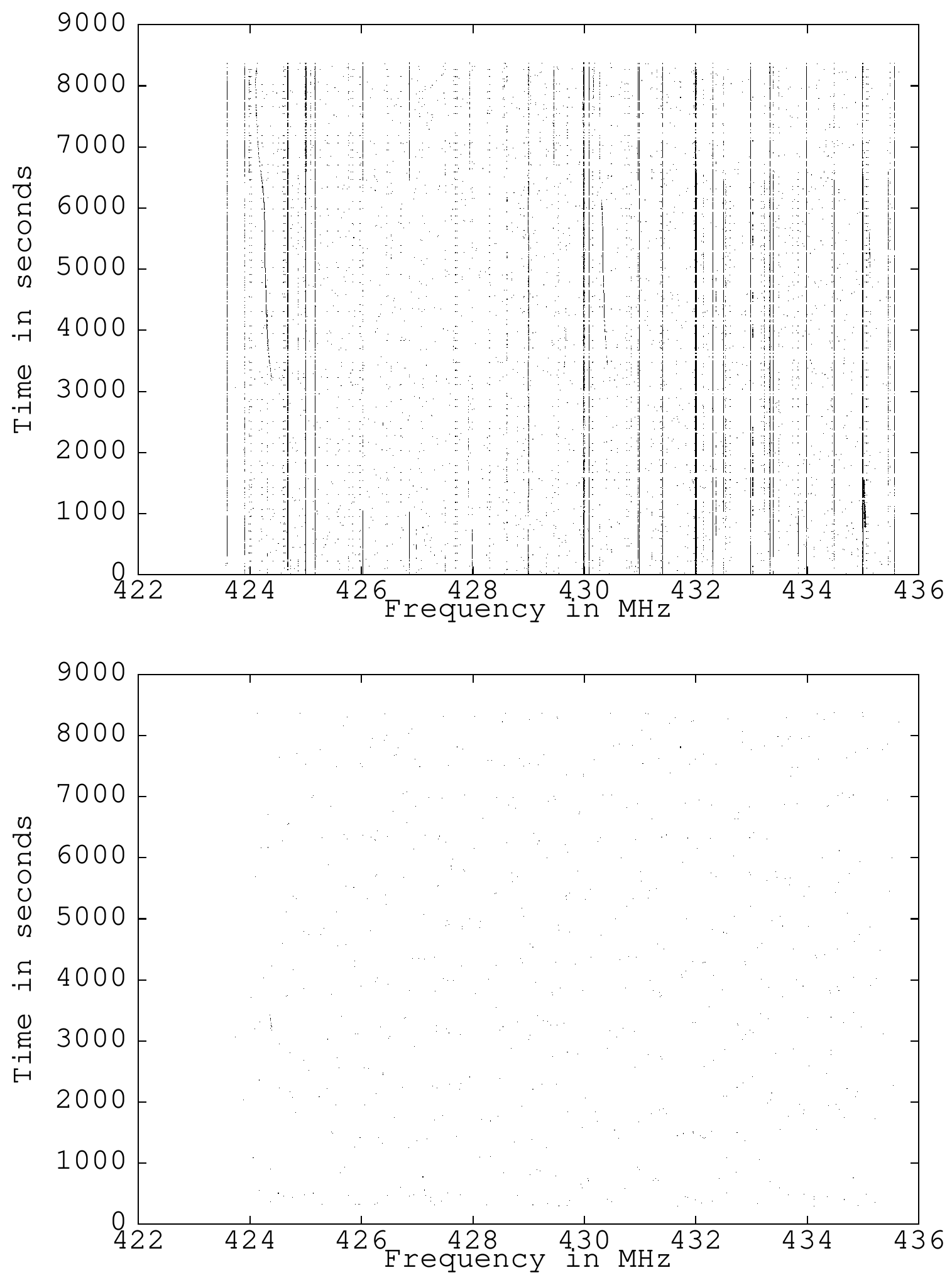}
\caption{Data from a period of severe interference (top). Each dot represents a
high power signal detection by the SERENDIP instrument. After RFI
filtering (bottom) the number of hit remaining approaches that expected
from a Poisson distribution, as would be expected in a
white-noise environment.\label{rfiexample}}
\end{center}
\end{figure}
}

\subsection{Pattern Detection}
A suite of pattern detection algorithms was
employed to detect signal recurrence and telescope beam pattern matching.
The primary pattern looked for was recurrence of a signal
in frequency over time and position.
All data were frequency corrected to the solar system barycenter.
Thus, only those signals that displayed the proper Doppler
drift relative to the solar system barycenter remained at a nearly
constant frequency. 
If a signal recurred within a defined frequency window,
it was tagged as a candidate.
We defined frequency windows of two different
widths in order to find two different classes of signals.

The first class (Class I) consists of signals with no frequency drift in the barycentric
frame.  If transmitted from a planet, such a signal must contain a Doppler
correction, and therefore would be a deliberately beamed message. 
For such signals, the chief cause of potential frequency drift arises 
from our lack of knowledge of the source's exact position on the sky.
We use the direction of the beam center to approximate the
source position, and obtain an approximate compensating 
frequency drift by projecting our antenna's diurnal and 
annual acceleration onto that nominal line of sight.  
Due to our beam's size (0.17 degrees), this drift rate uncertainty can 
shift an ideal signal by as much as 120 Hz.
Consequently, our compensated signal class defined a candidate
as any set of signals that recurred within a band whose width $f_\sigma=$ 120 Hz.

The second class (Class II) consists of signals containing significant Doppler
acceleration.
Radio leakage from a transmitter on a planet will significantly drift
in frequency over time spans of minutes to months.  
Here, signal sets were allowed a frequency deviation $f_\sigma=$ 50 kHz.
The width of this band was derived assuming a 
transmitter whose acceleration was as great as the Jovian cloud tops.

Multiple detections in a given transit were examined to determine
if the observed power as a function of time matched the telescope's 
Gaussian beam pattern.  Such a finding would suggest that the signal
emanated from a point source on the sky. 

\subsection{Candidate Merging and Ranking}
The output of SERENDIP's suite of pattern detection algorithms was a
collection of statistically interesting events.
These data sets were then examined to identify candidate locations for
signals of interest.
A candidate was defined as a one beamwidth area on the sky that
was identified by one or more pattern detection algorithms.  A
candidate is fully described by one or more records in the candidate database, 
each of which is the result of detection by a
single algorithm.  In addition, celestial objects such as nearby stars,
globular clusters, and known planetary systems are entered into the
candidate database to check for coincidence with candidates
identified by the other algorithms described herein.

Candidates are ranked by two methods.  First, each candidate record was given
an algorithm specific score.
Second, we asked how many algorithms detected the
candidate. Here we considered coincidence with an interesting celestial
object to be counted as an ``algorithm".

To determine the score in the case of multiple transit detections,
we calculated the relative
probability of our detections occurring from noise, given the number
of  detections in the candidate area, the number of times we have
observed the candidate area, the number of frequency bins searched by 
the algorithm, and the actual frequency separation of the detections.
The relative probability is given by
\begin{equation}
P = 
\frac{n^{n_d}_e}{{n_d}{!}{\left(\frac{F_{tot}}{F_{win}}\right)}^{({n_d}-1)}} 
\left[\frac{\Delta{f}+f_{\sigma}}{f_{\sigma}}\right]
\end{equation}
\noindent
where:

\noindent
$n_e$ is the total number of events logged at any  frequency in the candidate 
area,

\noindent
$n_d$ is the number of detections within $F_{win}$,

\noindent
$F_{win}$ is the frequency window searched for events,

\noindent
$F_{tot}$ is the total SERENDIP band observed,

\noindent
$\Delta{f}$ is the maximum frequency separation for the detection set,

\noindent
$f_{\sigma}$ is the expected frequency variance as explained in section 2.3.


\noindent
In searches for either Class I or Class II signals,
the lower the numerical rating in this
algorithm, the more promising the candidate.

To calculate the score in the case of 
beam pattern matching in a single transit across the telescope's
Gaussian beam, 
we first determined the best fit Gaussian to the data.
The relative score is then given by
\begin{equation}
P_{G} = A_{G}
\hspace{0.05in}
Q(\chi^2\hspace{0.05in}|\hspace{0.05in}\nu) 
\end{equation}
\noindent
where $A_{G}$ is the amplitude of the best fit Gaussian, and
$Q(\chi^2\hspace{0.05in}|\hspace{0.05in}\nu)$ is the
$\chi^2$ probability function.
In this case, the higher the numerical rating, the more
promising the candidate.

The data from the most promising candidates were independently scrutinized
for RFI contamination by three researchers
and those surviving this evaluation constitute
the list of most interesting candidates.
Our top candidates from our signal detection algorithms and their scores 
are given in Table 1.  

\ifpp{
\begin{deluxetable*}{ccccccccccc}
\tabletypesize{\scriptsize}
\tablenum{1}
\tablecaption{Most Interesting Candidates}
\tablehead{
\colhead{R.A.} & \colhead{Dec} & \colhead{Sky} & \colhead{Frequency} & \colhead{Max. Freq.} & \colhead{\# of} & \colhead{\# of} & \multicolumn{3}{c}{Score by} & \colhead{Coincident} \\
\colhead{} & \colhead{} & \colhead{Separation} & \colhead{} & \colhead{Separation\tablenotemark{1}} & \colhead{Obs.\tablenotemark{2}} & \colhead{Times} & \multicolumn{3}{c}{Algorithm\tablenotemark{3}} & \colhead{Object} \\
\colhead{(hours)} & \colhead{(deg)} & \colhead{(deg)} & \colhead{(MHz)} & \colhead{(Hz)} & \colhead{} & \colhead{Detected} & \colhead{1} & \colhead{2} & \colhead{3} & \colhead{} \\
}
\startdata
0.67 & 28.1 & 0.1 & 429.600 & 6 & 6 & 2 & 91.084 & & & GJ 1019 \\
0.96 & 7.8 & 0.1 & 426.839 & - & 1 & 1 & & & 43.17 & \\
2.13 & 26.9 & 0.1 & 431.959 & 17 & 20 & 3 & 1.801 & & & \\
4.59 & 27.0 & 0.1 & 434.603 & 30 & 13 & 3 & 0.632 & & & \\
5.13 & 2.1 & 0.2 & 435.009 & 46467 & 4 & 2 & & 79.371 & 22.68 & \\
8.15 & 9.0 & 0.1 & 435.089 & - & 1 & 1 & & & 26.93 & GJ 299 \\ 
10.23 & 18.5 & 0.2 & 430.059 & 79 & 34 & 3 & 20.525 & & & \\
14.32 & 25.9 & 0.2 & 427.350 & 72 & 8 & 3 & 0.250 & & & \\
23.09 & 26.8 & 0.1 & 429.973 & 27 & 21 & 3 & 2.529 & & & \\
\enddata
\tablenotetext{1}{Maximum frequency separation between multiple observations}
\tablenotetext{2}{Number of times SERENDIP III observed this point in the sky}
\tablenotetext{3}{Three different scoring algorithms:\\
Algorithm 1: Doppler compensated signal class (120Hz max frequency window) (see eq. 2); smaller is more remarkable\\
Algorithm 2: Non-Doppler compensated signal class (50kHz max frequency window) (see eq. 2); smaller is more remarkable\\
Algorithm 3: Gaussian beam fit (see eq. 3); larger is more remarkable}
\end{deluxetable*}
}

\subsection{Investigation of  Possible Signals from Nearby G-Dwarf Stars}

We investigated the data obtained from nearby G-dwarf stars that were
observed by chance in our sky survey with the rationale that these
were especially interesting candidates. In this effort, we used the
Center for Astrophysics list of G-dwarf stars within (roughly) 100pc of
the Sun (Latham 1999) as listed on their web site as of
August 1999\footnote{http://cfa-www.harvard.edu/$\sim$latham/gdwarf.html}.
We identified those
stars in this list visible with the Arecibo telescope (i.e. stars
between $-2^{\circ}$ and$+38^{\circ}$).   We then searched the SERENDIP
III position data set and identified stars in the list that fell within
the half power beam width of the Arecibo antenna during our
observations.

Of the  516 stars in the CFA list that  are observable with the Arecibo
telescope,  494 had been observed at least once, and 439 had been
observed multiple times. In Figure \ref{starsobs}\  we show a histogram of the number
of times these stars were observed.
\ifpp{
\begin{figure}[tbp]
\begin{center}
\includegraphics[width=\columnwidth,keepaspectratio=true]{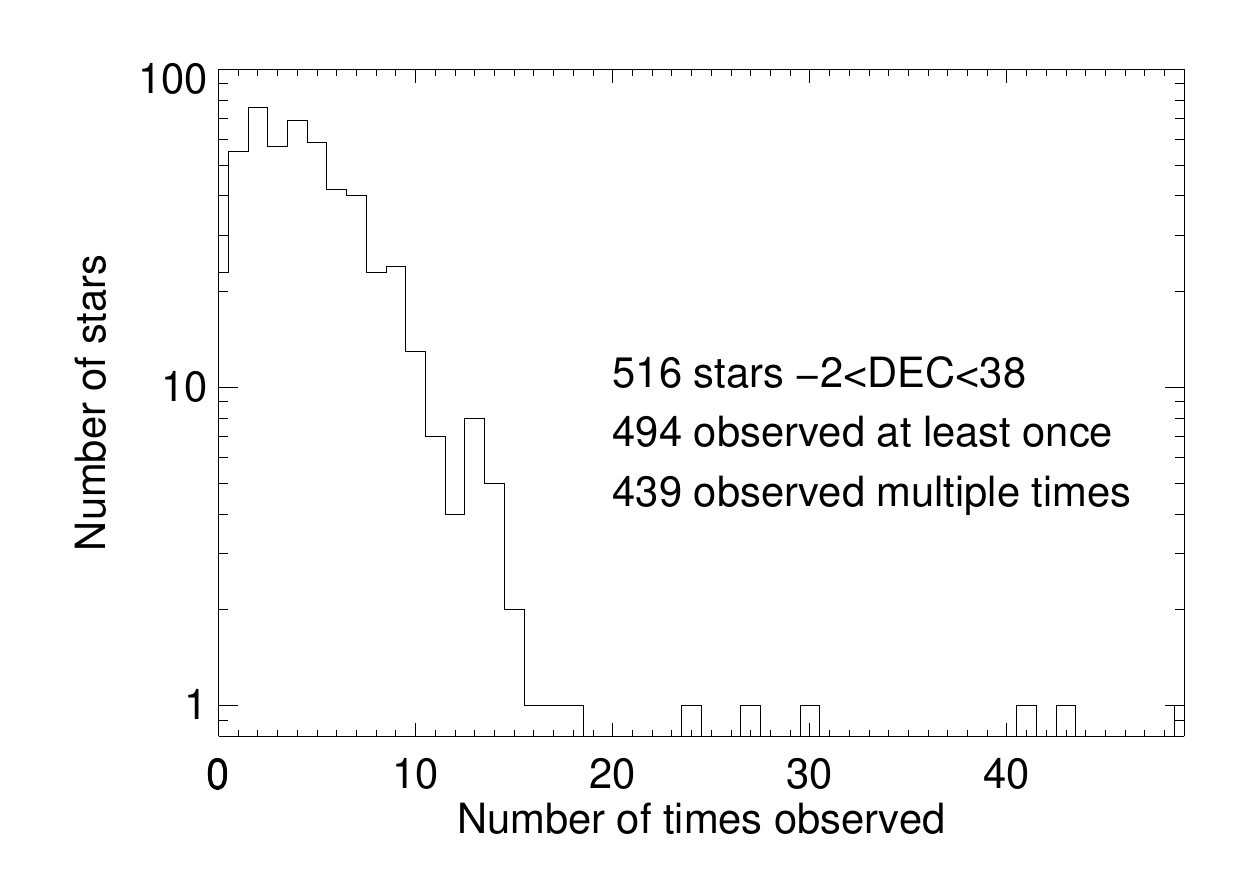}
\caption{A histogram of the number of separate times each
star in the CFA G-dwarf  subset  was observed in the  SERENDIP III
search.  No narrow band excess power beyond that expected for random
noise was detected from these stars.\label{starsobs}}
\end{center}
\end{figure}
}

We derived upper limits to the excess power from these stars by first
checking the most recent version of the Hipparcos catalogue for the
stellar distances.  The number of stars observed versus distance is
shown in Figure \ref{starsdist}. We then derived the upper limits to the source
stellar luminosity using the Hipparcos distances  and the observed
fluxes.  We note that the sensitivity limits are uncertain to a factor
of two because we used the average sensitivity of the antenna rather
than the sensitivity at the exact position of the star within the
beam.
\ifpp{
\begin{figure}[tbp]
\begin{center}
\includegraphics[width=\columnwidth,keepaspectratio=true]{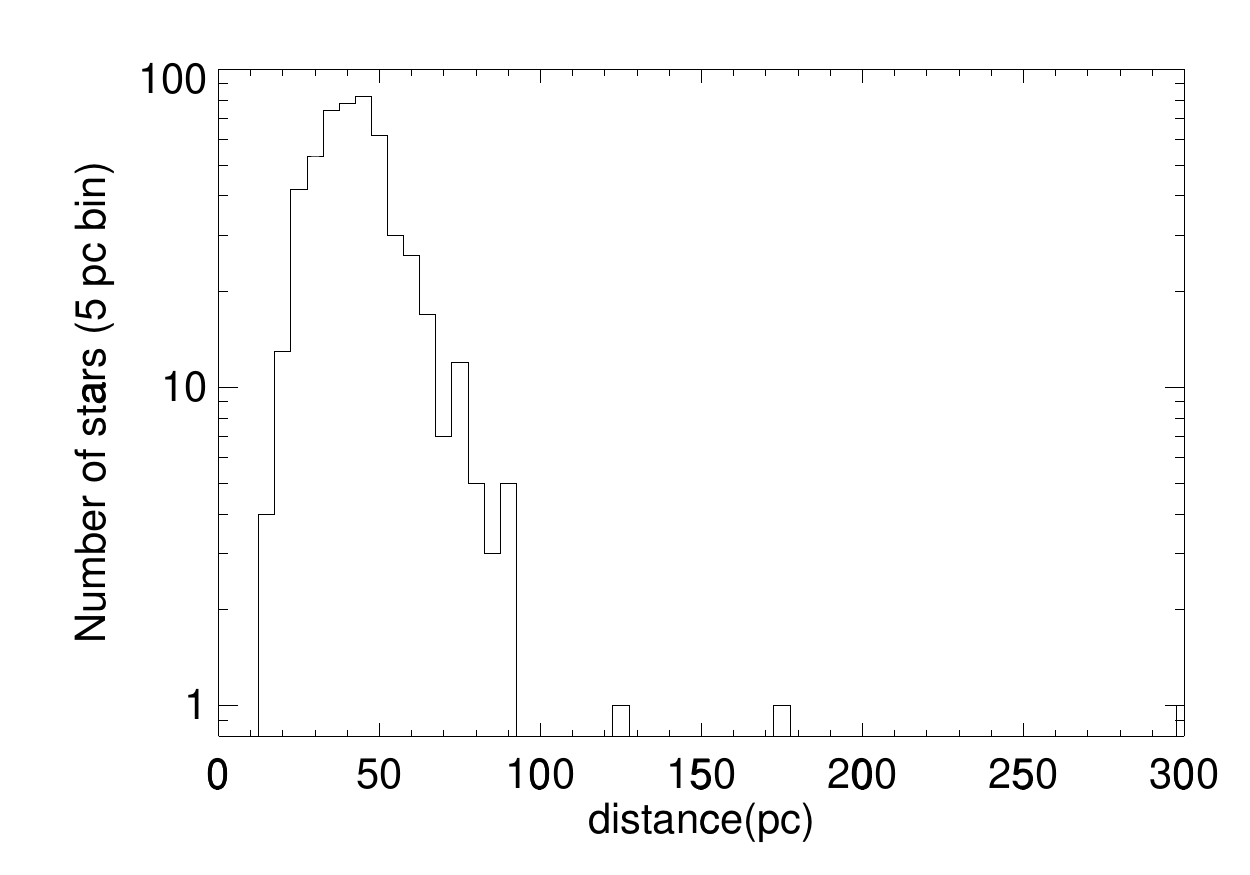}
\caption{A histogram showing the Hipparcos distance
distribution of stars in the CFA G-dwarf subset.\label{starsdist}}
\end{center}
\end{figure}
}

In Table 2 we provide the HD number of the observed stars, their
distances as obtained from the Hipparcos catalogue, and the upper
limits to the irradiated power at the star.

\section{Conclusions}

We have carried out an extensive search with the world's largest
telescope for evidence of radio emission produced by an extraterrestrial 
intelligence. We were able to carry out this search using this unique 
facility because of the non-intrusive
character of our observing program.  A major
challenge in our search (and in all other searches for
extra terrestrial intelligence) is the problem of false
signals produced as a result of human activity.  We
developed a variety of techniques to deal with this
problem and demonstrated their robustness.  This
shows that a non-intrusive collateral
data collection technique such as ours is viable.  

Given the extensive character of our search we found
many signals of potential interest.  A prioritization
scheme was developed to identify the most promising of
these signals. These were examined in more detail.  In 
the end, no extraterrestrial signals were identified. We 
do not find this surprising nor discouraging given our lack 
of knowledge as to appropriate source locations, frequencies and
time periods that an intentional extraterrestrial signal may be employing.

We are continuing our search. 

\acknowledgments 

We thank the Arecibo staff for their support in carrying out this
project which they did as an after-hours volunteer
project. 
The Arecibo Observatory is part of the National Astronomy and Ionosphere
Center which is operated by Cornell University under a Cooperative
Agreement with the National Science Foundation.
We acknowledge the efforts of the many volunteers who have
worked on SERENDIP over the years.  We acknowledge a generous gift by
Watson and Marilyn Alberts 
to carry out this analysis of the SERENDIP data.  This analysis has
also been supported by grants from the SETI Institute and the Planetary
Society.  The SERENDIP III project was supported by
significant contributions from Mrs. Elizabeth Bowyer, the Planetary Society,
and the Toshiba Corporation. Additional support was provided by
Intel, Xilinx, and the Friends of SERENDIP.

\ \\
\newpage
\clearpage
\LongTables
\begin{deluxetable}{crr}

\tabletypesize{\scriptsize}
\tablewidth{0pt}
\tablenum{2}
\tablecaption{Stars observed with SERENDIP III}
\tablehead{
\colhead{HD number} & \colhead{Distance} & \colhead{Upper Limit EIRP} \\
\colhead{} & \colhead{(pc)} & \colhead{(log$_{10}$(Watts))} }

\startdata
101 & 38 & 12.7 \\
377 & 40 & 12.8 \\
531E & 70 & 13.2 \\
531W & 70 & 13.2 \\
700 & 54 & 13.0 \\
1832 & 41 & 12.8 \\
4903 & 53 & 13.0 \\
5035 & 39 & 12.7 \\
6715 & 33 & 12.6 \\
7047A & 40 & 12.7 \\
8009A & 65 & 13.2 \\
8009B & 65 & 13.2 \\
8262 & 26 & 12.4 \\
8523 & 60 & 13.1 \\
8574 & 44 & 12.8 \\
8648 & 45 & 12.9 \\
9091 & 46 & 12.9 \\
9224 & 43 & 12.8 \\
9472 & 33 & 12.6 \\
9670 & 37 & 12.7 \\
9986 & 26 & 12.4 \\
10126 & 34 & 12.6 \\
10844 & 54 & 13.0 \\
11045 & 58 & 13.1 \\
11130 & 27 & 12.4 \\
11850 & 33 & 12.6 \\
12235 & 31 & 12.5 \\
12783 & 48 & 12.9 \\
13357A & 49 & 12.9 \\
13357B & 49 & 12.9 \\
13382 & 33 & 12.6 \\
13483 & 34 & 12.6 \\
13836 & 37 & 12.7 \\
13997 & 34 & 12.6 \\
14082B & 39 & 12.7 \\
14305 & 34 & 12.6 \\
14348 & 62 & 13.1 \\
14651 & 40 & 12.8 \\
14874 & 59 & 13.1 \\
15632 & 42 & 12.8 \\
16086 & 54 & 13.0 \\
16397 & 36 & 12.7 \\
17674 & 47 & 12.9 \\
17820 & 65 & 13.2 \\
18144 & 26 & 12.4 \\
18330 & 39 & 12.7 \\
18702 & 32 & 12.6 \\
18774 & 66 & 13.2 \\
19019 & 23 & 12.3 \\
19308A & 43 & 12.8 \\
19445 & 39 & 12.7 \\
19518 & 41 & 12.8 \\
19902 & 42 & 12.8 \\
19962 & 38 & 12.7 \\
20165 & 22 & 12.2 \\
20477 & 52 & 13.0 \\
21183 & 75 & 13.3 \\
21663A & 46 & 12.9 \\
21774 & 50 & 13.0 \\
22309 & 45 & 12.9 \\
23314 & 49 & 12.9 \\
24040 & 47 & 12.9 \\
24053 & 33 & 12.6 \\
24206 & 179 & 14.1 \\
24496 & 21 & 12.2 \\
24552 & 45 & 12.9 \\
24702 & 47 & 12.9 \\
25295 & 65 & 13.2 \\
25682 & 46 & 12.9 \\
25825 & 47 & 12.9 \\
26749 & 36 & 12.7 \\
26756 & 46 & 12.9 \\
26767 & 45 & 12.9 \\
26913 & 21 & 12.2 \\
26923 & 21 & 12.2 \\
27282 & 47 & 12.9 \\
27406 & 45 & 12.9 \\
27642 & 62 & 13.1 \\
27732 & 48 & 12.9 \\
27771 & 47 & 12.9 \\
27859 & 48 & 12.9 \\
28099 & 47 & 12.9 \\
28205 & 46 & 12.9 \\
28237 & 47 & 12.9 \\
28258 & 47 & 12.9 \\
28344 & 47 & 12.9 \\
28406B & 46 & 12.9 \\
28462 & 40 & 12.8 \\
28474 & 53 & 13.0 \\
28580 & 36 & 12.7 \\
28676 & 39 & 12.7 \\
28992 & 43 & 12.8 \\
29150 & 34 & 12.6 \\
29356 & 63 & 13.1 \\
29461 & 48 & 12.9 \\
30101E & 43 & 12.8 \\
30101W & 43 & 12.8 \\
30246 & 51 & 13.0 \\
30286 & 32 & 12.6 \\
30508 & 48 & 12.9 \\
30589 & 51 & 13.0 \\
30974 & 51 & 13.0 \\
31000 & 28 & 12.5 \\
31253 & 54 & 13.0 \\
31412 & 36 & 12.7 \\
31501 & 33 & 12.6 \\
31867 & 28 & 12.4 \\
32070 & 39 & 12.7 \\
32237 & 29 & 12.5 \\
32259 & 38 & 12.7 \\
32347 & 54 & 13.0 \\
33021A & 28 & 12.5 \\
33334NE & 47 & 12.9 \\
33334SW & 47 & 12.9 \\
33636 & 29 & 12.5 \\
33866 & 47 & 12.9 \\
34445 & 45 & 12.9 \\
34590 & 58 & 13.1 \\
34745 & 37 & 12.7 \\
35147 & 50 & 13.0 \\
35769 & 57 & 13.1 \\
36308 & 36 & 12.7 \\
36443 & 38 & 12.7 \\
36667 & 57 & 13.1 \\
37271 & 48 & 12.9 \\
37685 & 73 & 13.3 \\
37773 & 38 & 12.7 \\
38230A & 21 & 12.2 \\
39570A & 54 & 13.0 \\
39881A & 28 & 12.4 \\
40040 & 68 & 13.2 \\
40616 & 54 & 13.0 \\
41708 & 45 & 12.9 \\
42160 & 45 & 12.9 \\
42618 & 23 & 12.3 \\
43587A & 19 & 12.1 \\
43947 & 28 & 12.4 \\
44985 & 33 & 12.6 \\
45391 & 26 & 12.4 \\
45580 & 48 & 12.9 \\
45759 & 50 & 13.0 \\
46375 & 33 & 12.6 \\
47127 & 27 & 12.4 \\
48684 & 54 & 13.0 \\
50060 & 87 & 13.4 \\
51219 & 32 & 12.6 \\
51295 & 42 & 12.8 \\
52456 & 28 & 12.5 \\
53505 & 58 & 13.1 \\
53532 & 44 & 12.8 \\
54046 & 46 & 12.9 \\
54100 & 45 & 12.9 \\
54351 & 44 & 12.8 \\
54405 & 46 & 12.9 \\
54718 & 46 & 12.9 \\
55458A & 25 & 12.4 \\
55918 & 38 & 12.7 \\
56202 & 53 & 13.0 \\
56303 & 41 & 12.8 \\
56513 & 35 & 12.7 \\
58781 & 30 & 12.5 \\
58971 & 43 & 12.8 \\
59360 & 41 & 12.8 \\
59374 & 50 & 13.0 \\
60298 & 39 & 12.7 \\
62346 & 51 & 13.0 \\
63935 & 50 & 12.9 \\
64090 & 28 & 12.5 \\
64324 & 35 & 12.6 \\
65629 & 32 & 12.6 \\
66348A & 43 & 12.8 \\
66485 & 44 & 12.8 \\
66550 & 38 & 12.7 \\
68168 & 34 & 12.6 \\
68284 & 300 & 14.5 \\
69056 & 38 & 12.7 \\
70088 & 43 & 12.8 \\
70571A & 92 & 13.5 \\
70571B & 92 & 13.5 \\
72946 & 23 & 12.3 \\
73226 & 43 & 12.8 \\
73668A & 36 & 12.7 \\
74011 & 46 & 12.9 \\
74156 & 65 & 13.2 \\
74567 & 93 & 13.5 \\
75302 & 30 & 12.5 \\
76218 & 26 & 12.4 \\
76261 & 46 & 12.9 \\
76349 & 49 & 12.9 \\
76752 & 40 & 12.8 \\
76765 & 61 & 13.1 \\
76780 & 34 & 12.6 \\
77024A & 77 & 13.3 \\
77278 & 31 & 12.5 \\
77407 & 30 & 12.5 \\
78317 & 48 & 12.9 \\
78660 & 50 & 13.0 \\
79498A & 49 & 12.9 \\
79726 & 48 & 12.9 \\
80408 & 39 & 12.7 \\
80536 & 51 & 13.0 \\
80870 & 45 & 12.9 \\
81040 & 33 & 12.6 \\
82140 & 59 & 13.1 \\
82939 & 38 & 12.7 \\
83408 & 58 & 13.1 \\
84209 & 72 & 13.3 \\
84749 & 47 & 12.9 \\
85426 & 61 & 13.1 \\
85689 & 45 & 12.9 \\
86133A & 42 & 12.8 \\
86133B & 42 & 12.8 \\
86460 & 41 & 12.8 \\
86794 & 52 & 13.0 \\
87680 & 39 & 12.7 \\
88371 & 62 & 13.1 \\
88446 & 69 & 13.2 \\
88725 & 36 & 12.7 \\
89055 & 36 & 12.7 \\
89307 & 31 & 12.5 \\
89813 & 27 & 12.4 \\
90164 & 54 & 13.0 \\
90905 & 32 & 12.6 \\
91148 & 37 & 12.7 \\
91204 & 52 & 13.0 \\
93215 & 47 & 12.9 \\
94028 & 52 & 13.0 \\
94292 & 77 & 13.3 \\
94426 & 53 & 13.0 \\
95177 & 49 & 12.9 \\
95364 & 82 & 13.4 \\
95366 & 59 & 13.1 \\
95980 & 53 & 13.0 \\
96094 & 60 & 13.1 \\
96497 & 53 & 13.0 \\
96574 & 50 & 13.0 \\
96679 & 66 & 13.2 \\
96937 & 31 & 12.5 \\
98078 & 41 & 12.8 \\
98697 & 43 & 12.8 \\
99404 & 53 & 13.0 \\
99419 & 45 & 12.9 \\
99505 & 33 & 12.6 \\
100180A & 23 & 12.3 \\
100180B & 23 & 12.3 \\
101242 & 34 & 12.6 \\
101444 & 41 & 12.8 \\
103111 & 78 & 13.3 \\
103847 & 29 & 12.5 \\
104243 & 36 & 12.7 \\
104923 & 40 & 12.7 \\
104956 & 51 & 13.0 \\
105087 & 58 & 13.1 \\
105844 & 43 & 12.8 \\
106156 & 31 & 12.5 \\
106210 & 34 & 12.6 \\
106252 & 37 & 12.7 \\
106366 & 66 & 13.2 \\
106510 & 51 & 13.0 \\
107146 & 29 & 12.5 \\
107705A & 30 & 12.5 \\
107705B & 30 & 12.5 \\
108653 & 51 & 13.0 \\
108874 & 69 & 13.2 \\
109628A & 84 & 13.4 \\
109628B & 84 & 13.4 \\
111398A & 36 & 12.7 \\
111470 & 56 & 13.1 \\
112001 & 55 & 13.0 \\
112257 & 42 & 12.8 \\
112735 & 58 & 13.1 \\
112756 & 53 & 13.0 \\
112959A & 59 & 13.1 \\
113319 & 31 & 12.5 \\
114060A & 37 & 12.7 \\
114060B & 37 & 12.7 \\
114174 & 26 & 12.4 \\
114606 & 61 & 13.1 \\
115231 & 53 & 13.0 \\
115273 & 49 & 12.9 \\
115349 & 45 & 12.9 \\
115382 & 91 & 13.5 \\
115519 & 61 & 13.1 \\
115755 & 34 & 12.6 \\
115762A & 60 & 13.1 \\
116442 & 16 & 12.0 \\
116443 & 17 & 12.0 \\
116497 & 49 & 12.9 \\
117858 & 60 & 13.1 \\
118659 & 53 & 13.0 \\
119054 & 50 & 12.9 \\
119056 & 61 & 13.1 \\
119550 & 60 & 13.1 \\
120553 & 44 & 12.8 \\
121298NW & 75 & 13.3 \\
121298SE & 75 & 13.3 \\
121320 & 33 & 12.6 \\
122518 & 55 & 13.0 \\
122652 & 37 & 12.7 \\
123033B & 44 & 12.8 \\
124019 & 49 & 12.9 \\
124677A & 36 & 12.7 \\
125056 & 42 & 12.8 \\
126246A & 36 & 12.7 \\
126246B & 36 & 12.7 \\
126512 & 47 & 12.9 \\
126583 & 34 & 12.6 \\
126961 & 41 & 12.8 \\
127825 & 58 & 13.1 \\
128219 & 68 & 13.2 \\
129209 & 50 & 13.0 \\
129413A & 40 & 12.8 \\
129814 & 42 & 12.8 \\
130268 & 68 & 13.2 \\
131179 & 39 & 12.7 \\
132973 & 75 & 13.3 \\
133161 & 35 & 12.7 \\
134066A & 32 & 12.6 \\
134066B & 32 & 12.6 \\
135101A & 28 & 12.5 \\
135101B & 28 & 12.5 \\
135792A & 43 & 12.8 \\
136925 & 46 & 12.9 \\
138246 & 62 & 13.1 \\
138573 & 31 & 12.5 \\
138919 & 41 & 12.8 \\
139018 & 79 & 13.3 \\
139324 & 53 & 13.0 \\
139457 & 47 & 12.9 \\
139839 & 65 & 13.2 \\
140209 & 65 & 13.2 \\
140233 & 78 & 13.3 \\
140324 & 55 & 13.0 \\
140514 & 83 & 13.4 \\
140750 & 68 & 13.2 \\
141272 & 21 & 12.2 \\
141529 & 55 & 13.0 \\
142093 & 31 & 12.5 \\
142229 & 41 & 12.8 \\
142637 & 66 & 13.2 \\
143291 & 26 & 12.4 \\
144873 & 47 & 12.9 \\
145229 & 33 & 12.6 \\
145729 & 45 & 12.9 \\
146588 & 45 & 12.9 \\
146644 & 61 & 13.1 \\
147044 & 36 & 12.7 \\
147528 & 51 & 13.0 \\
147750 & 40 & 12.8 \\
148530 & 47 & 12.9 \\
148816 & 41 & 12.8 \\
149028 & 48 & 12.9 \\
149380 & 94 & 13.5 \\
149890 & 39 & 12.7 \\
150554A & 45 & 12.9 \\
150828B & 63 & 13.2 \\
150933A & 44 & 12.8 \\
152264 & 64 & 13.2 \\
153627 & 43 & 12.8 \\
153701 & 37 & 12.7 \\
154417 & 20 & 12.2 \\
154656 & 42 & 12.8 \\
154931 & 55 & 13.0 \\
155060 & 36 & 12.7 \\
155193 & 61 & 13.1 \\
155358 & 43 & 12.8 \\
155423 & 44 & 12.8 \\
156146 & 76 & 13.3 \\
156893 & 68 & 13.2 \\
156968 & 55 & 13.0 \\
157089 & 39 & 12.7 \\
157637 & 41 & 12.8 \\
158226A & 69 & 13.2 \\
158331 & 51 & 13.0 \\
158332 & 30 & 12.5 \\
159909 & 37 & 12.7 \\
160013 & 42 & 12.8 \\
161728 & 63 & 13.2 \\
161848 & 38 & 12.7 \\
162209 & 53 & 13.0 \\
163609A & 50 & 12.9 \\
164595 & 29 & 12.5 \\
165173 & 33 & 12.6 \\
165476 & 45 & 12.9 \\
165672 & 44 & 12.8 \\
166301 & 35 & 12.6 \\
167081A & 49 & 12.9 \\
169359 & 60 & 13.1 \\
169748A & 52 & 13.0 \\
169822 & 27 & 12.4 \\
169889 & 38 & 12.7 \\
170294 & 58 & 13.1 \\
170469 & 65 & 13.2 \\
171009 & 70 & 13.2 \\
171067 & 25 & 12.4 \\
171620 & 52 & 13.0 \\
172310 & 36 & 12.7 \\
172649 & 47 & 12.9 \\
172867 & 66 & 13.2 \\
173548A & 52 & 13.0 \\
173548B & 52 & 13.0 \\
174457 & 55 & 13.0 \\
174719 & 28 & 12.5 \\
175726 & 27 & 12.4 \\
177305 & 43 & 12.8 \\
178911B & 49 & 12.9 \\
180684 & 59 & 13.1 \\
181047A & 50 & 13.0 \\
182619 & 33 & 12.6 \\
182758 & 66 & 13.2 \\
183341 & 45 & 12.9 \\
183970 & 50 & 13.0 \\
184403 & 44 & 12.8 \\
184592 & 42 & 12.8 \\
186704A & 30 & 12.5 \\
187123 & 48 & 12.9 \\
187548 & 45 & 12.9 \\
187882 & 76 & 13.3 \\
187897 & 33 & 12.6 \\
187923A & 28 & 12.4 \\
188015 & 53 & 13.0 \\
188510 & 39 & 12.7 \\
189067 & 43 & 12.8 \\
189087 & 25 & 12.4 \\
190404 & 16 & 11.9 \\
190516A & 44 & 12.8 \\
190594 & 43 & 12.8 \\
190605 & 44 & 12.8 \\
190609 & 44 & 12.8 \\
191672 & 69 & 13.2 \\
192367 & 129 & 13.8 \\
195019A & 37 & 12.7 \\
195034 & 28 & 12.5 \\
196201 & 38 & 12.7 \\
196885A & 33 & 12.6 \\
198089 & 39 & 12.7 \\
198416 & 60 & 13.1 \\
200466E & 44 & 12.8 \\
200466W & 44 & 12.8 \\
200565 & 64 & 13.2 \\
200746 & 44 & 12.8 \\
201219 & 36 & 12.7 \\
201891 & 35 & 12.7 \\
202072 & 49 & 12.9 \\
202108 & 27 & 12.4 \\
203030 & 41 & 12.8 \\
204277 & 34 & 12.6 \\
204712 & 61 & 13.1 \\
205702 & 57 & 13.1 \\
206332 & 50 & 13.0 \\
206374 & 27 & 12.4 \\
206387A & 55 & 13.0 \\
206658 & 48 & 12.9 \\
207740 & 49 & 12.9 \\
209262A & 46 & 12.9 \\
209458 & 47 & 12.9 \\
209858 & 55 & 13.0 \\
209875 & 51 & 13.0 \\
210388 & 43 & 12.8 \\
210460 & 56 & 13.0 \\
210462A & 54 & 13.0 \\
210483 & 49 & 12.9 \\
210553 & 45 & 12.9 \\
211476 & 31 & 12.5 \\
211786 & 42 & 12.8 \\
212291 & 32 & 12.6 \\
212858 & 55 & 13.0 \\
214059 & 80 & 13.4 \\
214435 & 49 & 12.9 \\
214560 & 55 & 13.0 \\
215257 & 42 & 12.8 \\
215274 & 45 & 12.9 \\
216625 & 44 & 12.8 \\
216631 & 43 & 12.8 \\
217165 & 44 & 12.8 \\
218133 & 38 & 12.7 \\
218172 & 74 & 13.3 \\
218261 & 28 & 12.5 \\
219172 & 46 & 12.9 \\
220008 & 87 & 13.4 \\
220077 & 77 & 13.3 \\
220255 & 52 & 13.0 \\
220334B & 37 & 12.7 \\
220773 & 48 & 12.9 \\
221477 & 59 & 13.1 \\
221822 & 39 & 12.7 \\
221851 & 23 & 12.3 \\
221876 & 75 & 13.3 \\
222033 & 50 & 13.0 \\
222941A & 46 & 12.9 \\
223061 & 45 & 12.9 \\
223238 & 47 & 12.9 \\
224156 & 29 & 12.5 \\
225261 & 26 & 12.4 \\
\end{deluxetable} 
\ \\
\newpage
\clearpage
\end{document}